\documentstyle[pre,aps,epsfig]{revtex}

\newcommand{\gtapprox}{\hbox{$>\mkern-19mu\lower4pt\hbox{$\sim$}$}}
\newcommand{\ltapprox}{\hbox{$<\mkern-19mu\lower4pt\hbox{$\sim$}$}}

\begin{document}


\title{Parallel Excluded Volume Tempering for Polymer Melts}
\author{Alex Bunker and Burkhard D{\"u}nweg} 
\address{Max Planck Institute for Polymer Research, Ackermannweg 10,
D-55128 Mainz, Germany}
\date{\today}
\maketitle

\begin{abstract}
  We have developed a technique to accelerate the
  acquisition of effectively uncorrelated configurations for
  off-lattice models of dense polymer melts which makes use of both
  parallel tempering and large scale Monte Carlo moves. 
  The method is based upon
  simulating a set of systems in parallel, each of which has a
  slightly different repulsive core potential, such that a
  thermodynamic path from full excluded volume to an ideal gas of
  random walks is generated. While each system is run with standard
  stochastic dynamics, resulting in an NVT ensemble, we implement 
  the parallel tempering through stochastic
  swaps between the configurations of adjacent potentials, and the
  large scale Monte Carlo moves through attempted pivot and
  translation moves which reach a realistic acceptance probability
  as the limit of the ideal gas of random walks is approached. 
  Compared to pure stochastic dynamics,
  this results in an increased efficiency even for a system of chains
  as short as $N = 60$ monomers, however at this chain length the
  large scale Monte Carlo moves were ineffective. For even longer
  chains
  the speedup becomes substantial, as observed from preliminary data
  for $N = 200$.
  While the previously established end-bridging
  algorithm relaxes the end-to-end autocorrelation function
  more quickly, it does so at the price of an artificial
  polydispersity, which the current method does not exhibit. 
  The end-to-end autocorrelation function is however no longer
  the slowest mode for the end-bridging algorithm so whether
  the algorithm is superior even for the case of polydispersity is unclear.
\end{abstract}


\section{Introduction}

Computer simulations of dense polymer systems which make use of
off-lattice models have been successful in the determination of both
static properties, such as phase equilibria \cite{grest_1996,%
  escobedo_1996,escobedo_1997,escobedo_1997_1} or rubber elasticity
\cite{everaers_1999}, and dynamic properties, such as the details of
single-chain and collective relaxation \cite{kremer_1990,%
  kroeger_1993,kopf_1997,paul_1998}. For a melt of polymers of
length $N$, the relaxation time $\tau$, 
the time taken for the polymer to assume a
new configuration, scales as
$\tau \propto N^2$ for small $N$ (Rouse dynamics), while for
larger $N$ reptation behavior, $\tau \propto N^3$
\cite{kremer_1990,doi_1986} sets in. If the computer
simulation of a polymer melt is performed in such a way that
the dynamic properties are realistically reproduced
then this scaling will be directly related to
the computational effort needed to effectively sample phase space
and obtain meaningful results for the static properties
\cite{duenweg_1998}. As a result progress in the simulation of systems
involving polymers with large $N$ has been severely hampered.

A modern trend in Monte Carlo simulations in statistical physics is to
strictly distinguish between simulations which only aim to determine
the static properties of a given system, and those which also set out
to determine the dynamic behavior. In the latter case one has to
follow the natural motion of the system confined to local dynamics
constrained by topology and/or barriers.  If one is however only
interested in generating uncorrelated configurations as quickly as
possible, one can use an artificial dynamics which is able to reach
new effectively uncorrelated configurations much more quickly than the
physical dynamics would allow.

The Rouse scaling law $\tau \propto N^2$
is a direct consequence of only allowing local motions to
occur. It is independent of any constraint to motion and holds even
for phantom chains with no interaction whatsoever except connectivity.
Clearly, violating locality is an important step if one wishes to
accelerate the acquisition of uncorrelated configurations.
Particularly successful examples of schemes which achieve this include
cluster algorithms \cite{swendsen_1987} for critical phenomena, and
the pivot algorithm \cite{madras_1988,sokal_1995} for isolated polymer
chains, which collectively rotates a large part of the chain at once,
thus allowing one to study the static properties for
$N = 10^5$ and above \cite{sokal_1995}.  

In a dense polymer system, however, such an approach will clearly fail,
since practically any attempted large scale move will be rejected
due to overlap with other monomers. The effective constraints to
motion which cause these techniques to fail are of physical importance 
--- 
this is the mechanism which, for sufficiently long chains,
gives rise to the onset of the
considerably slower reptation dynamics.
As a result previous attempts
to speed up simulations of polymer melts by {\em only} lifting
locality are unable to alleviate the problem.  For example, the
continuum configurational biased Monte Carlo method (CCB,
\cite{frenkel_1996}) and its variants \cite{consta_1999} remove a
chain (partly), and attempt to regrow it into the existing matrix.
This can in principle be seen as a non-local approach like the pivot
algorithm, however, in a simulation of a dense polymer melt the chain
will grow preferentially back into the cavity from which it was
previously removed. This effect becomes more pronounced with
increasing chain length.

A simulation algorithm geared at only generating uncorrelated
equilibrium configurations should thus not only find a way to violate
locality but also the constraints resulting from the topology of the
system and/or barriers. Fortunately, techniques have been developed to
achieve this.  The multicanonical ensemble and its variants (also
called ``umbrella sampling'', ``entropic sampling'' or ``$1/k$ sampling'')
\cite{valleau_1977,torrie_1977,berg_1992,berg_1992_1,lee_1993,%
hesselbo_1995} try to identify barriers and then introduce a
suitable bias in order to remove them (i.~e. to allow the system to
easily enter these unfavorable states). Simulated tempering
(also called ``expanded ensemble'') \cite{marinari_1992,%
lyubartsev_1992} tries to systematically soften the constraints to
motion by giving the system access to different parameter values where
the barriers are weaker. 

While previous implementations of simulated tempering have mostly
limited themselves to using an intensive variable of the ensemble,
such as temperature or chemical potential, as the control parameter,
an admittedly less intuitively obvious choice can however be made.
The control parameter can instead originate from a term within the
Hamiltonian of the ensemble itself.  For a model of a polymer melt the
strength of the excluded volume interaction is a particularly useful
parameter, since it directly generates the topological constraints
which ultimately give rise to reptation-like slowing down.  This idea
has already successfully been applied to lattice polymers for
measuring chemical potentials \cite{mueller_1993,wilding_1994} and
within the framework of a multicanonical ensemble \cite{iba_1998},
while for continuum polymers it has so far only been used in an {\em ad
hoc} fashion for equilibration purposes
\cite{kremer_1990,kopf_1997}.

The present investigation is a first attempt to combine the virtues of
{\em both} non-local updates {\em and} constraint removal for
continuum models of dense polymer melts. The former is done via pivot
moves \cite{madras_1988,sokal_1995}, while the latter is done via
parallel tempering in the strength of the excluded volume interaction.
Parallel tempering, also called ``multiple Markov chains''
\cite{geyer_1991,orlandini_1998} or ``exchange Monte Carlo''
\cite{hukushima_1996}, is very similar in spirit to simulated
tempering \cite{marinari_1992,lyubartsev_1992}, but offers a number of
both conceptual and technical advantages. 

Both approaches are based on
studying a whole family of Hamiltonians ${\cal H}_i$, $i = 1, \ldots,
n$, each of which defines a standard Boltzmann weight $\exp \left( -
  {\cal H}_i \right)$, where, for convenience, the temperature has
been absorbed into the definition of the Hamiltonian. This family of
Hamiltonians will form a sequence in the one dimensional space of
the control parameter. Along this line, the Hamiltonians must be
located close enough
to each other, such that the distribution of
equilibrium states resulting from the Boltzmann weight $\exp \left( -
  {\cal H}_i \right)$ has significant overlap with the
distributions given by the Boltzmann weights $\exp \left( - {\cal
    H}_{i - 1} \right)$ and $\exp \left( - {\cal H}_{i + 1} \right)$.
A typical configuration for Hamiltonian ${\cal H}_i$ should
be within the thermal fluctuations for both Hamiltonians, ${\cal H}_{i - 1}$ 
and ${\cal H}_{i + 1}$.
As system size increases the distributions become sharper and sharper.
As a result more and more Hamiltonians will be required for this condition to
still be satisfied. One would thus in principle
like to study a system which is as small as possible.

If we denote the control parameter by $\phi$,
the above condition can be expressed as follows:
For the averages of a given extensive variable
$A$ in two adjacent ensembles characterized by $\phi$ and $\phi +
\Delta \phi$ the relation
\begin{equation}
\left\vert 
\left< A \right>_{\phi + \Delta \phi} - \left< A \right>_\phi
\right\vert \approx
\left\vert \frac{\partial \left< A \right>}{\partial \phi} \right\vert
\left\vert \Delta \phi \right\vert \ltapprox
\left( \left< A^2 \right>_\phi - \left< A \right>_\phi^2 \right)^{1/2}
\end{equation}
should hold. Since
\begin{equation}
\left( \left< A^2 \right>_\phi - \left< A \right>_\phi^2 \right)^{1/2} 
\propto V^{1/2}
\end{equation}
for reasons of Gaussian statistics and $\partial A /
\partial \phi \propto V$, one finds $\Delta \phi \propto V^{-1/2}$ or
$n$, the number of Hamiltonians in the sequence, $\propto V^{1/2}$. 

Given a family of Hamiltonians with the above condition satisfied, the
tempering procedure consists of allowing a given system to make
stochastic switches to neighboring Hamiltonians on the sequence in
parameter space at fixed system configuration.  Ideally, this results
in a diffusion process with respect to the Hamiltonians. In
particular, a configuration which was originally subject to a ``hard''
Hamiltonian (with constraints) can diffuse to a ``soft'' Hamiltonian
(without), relax there quickly, and return to the original hard
Hamiltonian. This should, hopefully, accelerate the rate at which the
system traverses phase space.

For a dense three-dimensional melt of flexible polymers one expects
the following scaling: The time to diffuse along the path of
Hamiltonians and back is proportional to $n^2 \propto V$. Assuming
that the soft Hamiltonian does not provide any constraints, and that a
suitable algorithm is able to generate there a completely new
configuration in practically zero relaxation time, one finds
altogether $\tau \propto V$. Furthermore, the smallest system one can
study is given by equating the linear box size to the mean end-to-end
distance $R \propto N^{1/2}$ (in a melt the conformations are random
walks \cite{doi_1986}). Thus $\tau \propto V \propto N^{3/2}$, which
is somewhat better than plain Rouse relaxation, $\tau \propto N^2$,
and considerably faster than reptation, $\tau \propto N^3$.
Nevertheless, it should be noted that the well-known slithering-snake
algorithm \cite{sokal_1995} scales as $\tau \propto N^{\approx 1}$,
i.~e. is expected to be asymptotically even better than our procedure.
For very dense systems the prefactor in this law will however be
large, due to small acceptance rates of the slithering-snake moves,
such that one might need unrealistically long chains in order to
actually observe the superiority. Where we expect the biggest payoff
for our algorithm, however, is in systems where the (true physical)
dynamics is governed by an activated process, such as star polymers
\cite{mcleish_1999}, where
\begin{equation}
\tau \propto \exp \left( {\rm const.} N \right) ,
\end{equation}
and for which the slithering-snake algorithm is not applicable.
Regardless of these considerations, our first tests have deliberately
focused on melts of linear chains, since this is the system which is
characterized best with respect to both statics and dynamics.

The difference between simulated tempering and parallel tempering
originates in how this idea is put into practice. Standard
simulated tempering \cite{marinari_1992,lyubartsev_1992} considers
only one system, whose configurations we denote by $\vec x$, and
simply adds the parameter $\phi$ as an additional degree of freedom,
which is treated via a standard Monte Carlo algorithm in that expanded
state space. This procedure is governed by the Hamiltonian ${\cal H}
(\phi,\vec x) - \eta(\phi)$, where $\eta$ is a suitable pre-weighting
factor, to be determined self-consistently in order to prevent the
simulation from getting trapped in the softest Hamiltonian. The
partition function of the resulting expanded ensemble is given by 
\begin{equation}
Z = \sum_i \exp \left( \eta_i \right)
    \int d \vec x \exp \left( - {\cal H}_i \right)
  = \sum_i \exp \left( \eta_i - F_i \right),
\end{equation}
where $F$ is the free energy 
(temperature is again absorbed in the definition). 
Since the arguments of the exponentials
are extensive, the sum will always be strongly dominated by the
largest term, unless all of them are practically identical. This means 
that unless $\eta_i \approx F_i$ $\forall$ $i$ the system will not be able
to traverse the full extent of the available parameter space as
one or more parameter values will become highly improbable.  

In parallel tempering $n$ systems are run in parallel, 
each of which is assigned one of the Hamiltonians
${\cal H}_i$. Diffusion in Hamiltonian space is then facilitated by
simple swaps of the configurations of adjacent Hamiltonians.  Since
each Hamiltonian will always be occupied, there is no problem of the
simulation not visiting any particular ``unfavorable'' Hamiltonian,
and thus it is no longer necessary to determine pre-weighting factors. 
Furthermore, the
scaling considerations from above remain valid; the increased CPU
effort by a factor of $n$ is rewarded by the fact that we now have $n$
random walkers available to produce data. The series of $n$ systems
can be seen as one extended ensemble with the partition function
\begin{eqnarray}
Z & = & \nonumber
\int d \vec x_1 \ldots d \vec x_n \frac{1}{n !} \sum_p \\
  &   & \nonumber
\exp \left( - {\cal H}_{p(1)} (\vec x_1) \right) \ldots
\exp \left( - {\cal H}_{p(n)} (\vec x_n) \right) \\
  & = & \prod_i Z_i ,
\end{eqnarray}
where $p$ denotes the possible permutations of the index set $i = 1,
\ldots, n$, and we have made use of the arbitrariness in labeling.
Thus the method just simulates $n$ statistically independent systems.

The detailed balance condition for the swap is derived in a 
straightforward manner: If we denote two systems in which we attempt to
switch the Hamiltonians by $\vec{x}$ (governed initially by
Hamiltonian ${\cal H}_1$) and $\vec{y}$ (governed initially by
Hamiltonian ${\cal H}_2$), then the transition probabilities $w$
must satisfy
\begin{eqnarray}
& &
\frac
{ w \left( \left( \vec x, \vec y \right) \to
           \left( \vec y, \vec x \right) \right) }
{ w \left( \left( \vec y, \vec x \right) \to
           \left( \vec x, \vec y \right) \right) } =
\frac
{ P_{eq} (\vec y, \vec x) }
{ P_{eq} (\vec x, \vec y) } \\ & = & \nonumber
\exp \left(
- {\cal H}_1 (\vec y)
- {\cal H}_2 (\vec x)
+ {\cal H}_1 (\vec x)
+ {\cal H}_2 (\vec y) \right) =: B
\end{eqnarray}
(the partition functions cancel out in the ratio of equilibrium
distributions). Using the standard Metropolis rule, the attempted swap
$(\vec x, \vec y) \to (\vec y, \vec x)$ is accepted with probability
$\min (1,B)$.

Given the simplicity of the method, and its potential, one should
expect that its popularity will increase substantially in the future.
So far, its use has not been very widespread, partly due to the fact
that access to massively parallel computing facilities (for which the
approach is ideally suited) is still somewhat limited. Applications
up to now have included spin glasses \cite{hukushima_1996},
liquid-vapor phase coexistence \cite{yan_1999}, and several studies
on the theta collapse of single polymer chains, and related issues,
where always the temperature was used as the control parameter
\cite{orlandini_1998,tesi_1996,nidras_1997,hansmann_1997,%
  orlandini_1998_1,rensburg_1999,irback_1999}. The fact that the
strength of the excluded volume interaction could be used as a
parameter in parallel tempering is mentioned in Ref.
\onlinecite{iba_1998}; however, no actual run data were presented.

The remainder of this paper is organized as follows: In Sec.
\ref{modelsec} we describe the details of model and algorithm. Section
\ref{kgmodelsec} defines the standard Kremer-Grest model of a polymer
melt \cite{kremer_1990}, and its simulation by means of stochastic
Langevin dynamics. Section \ref{partempsec} then describes the most
important ingredients of our parallel tempering procedure, which is
based upon altering the functional form of the repulsive core
potential and replacing it by a non-divergent ``soft-core'' potential,
until the limit of phantom chains is reached. This allows the chains
to pass through each other, thus eliminating the slow reptation
dynamics. When the repulsive core potential is soft enough we will be
able to perform pivot and whole polymer translation moves in the melt
for which $\tau \propto N^{\approx 0}$, as described in more detail in
Sec. \ref{pivotsec}. We also compare with end-bridging \cite{pant_1995,%
  mavrantzas_1999}, a very fast Monte Carlo algorithm which however
does not conserve the chain lengths, and whose basic features are
outlined in Sec. \ref{endbridgesec}. Section \ref{resultsec} reports
our numerical results. Section \ref{construsec} describes how we
found the parameters for our procedure, while important time
correlation functions to measure the efficiency of our algorithm
are defined and presented in Sec. \ref{corrfuncsec}, resulting
in our conclusion (Sec. \ref{conclusec}).

\section{Model and Algorithm}
\label{modelsec}

\subsection{Kremer-Grest Model and Langevin Dynamics}
\label{kgmodelsec}

The Kremer-Grest model \cite{kremer_1990} is one of several
off-lattice models for polymer melts which are commonly known as
``bead-spring'' models.  All particles have purely repulsive
Lennard-Jones cores of the form
\begin{eqnarray}
U_{LJ}(r) &=&
4\epsilon
\left[\left(\frac{\sigma}{r}\right)^{12}-
\left(\frac{\sigma}{r}\right)^{6}+\frac{1}{4}\right]
\hspace*{0.4in} r\le 2^{1/6}\sigma, \nonumber    \\
U_{LJ}(r) &=& 0 
\hspace*{1.85in} r \ge 2^{1/6}\sigma,
\end{eqnarray}
where $\epsilon$ and $\sigma$, as well as the bead mass, are set to
unity such that time is in Lennard-Jones units. The FENE
attraction between the neighboring monomers on the chains is given by
\begin{equation}
U_{ch}(r) = -\frac{k}{2}R_0^2\ln\left(1-\frac{r^2}{R_0^2}\right),
\end{equation}
where $R_0 = 1.5$ is the maximum extension of the nonlinear spring,
and $k = 30$ is the spring constant. The spring constant is set to be strong
enough to prohibit two polymer chains from crossing each other. We
consider a system of $M$ chains of length $N$ in a cubic box with
periodic boundary conditions at constant volume with density $\rho =
0.85$.

We have simulated an NVT ensemble of this system through the use of
Langevin (stochastic) dynamics \cite{kremer_1990,duenweg_1998}, fixing
the temperature at $k_B T = 1.0$ where $k_B$ denotes Boltzmann's
constant. This involves the addition of a random force and a friction
term, resulting in the following equations of motion in terms of
particle positions $\vec r_i$ and momenta $\vec p_i$:
\begin{eqnarray}
\dot{\vec p_i} & = & \vec F_i - \frac{\gamma}{m_i} \vec p_i +
\vec f_i , \nonumber\\
\dot{\vec r_i} & = & \frac{\vec p_i}{m_i} ,
\end{eqnarray}
where $\vec F_i$ is the force due to the interactions with other
monomers, $m_i$ the particle mass, $\gamma$ the friction constant, and
$\vec f_i$ the stochastic force which satisfies the standard
fluctuation-dissipation relation
\begin{equation}
\left< f_{i \alpha} (t) f_{j \beta} (t^\prime) \right>
= 2 \gamma k_B T \delta_{ij} \delta_{\alpha \beta} 
\delta (t - t^\prime)
\end{equation}
(i.~e. uncorrelated with respect to both particle indices $i$ and
Cartesian indices $\alpha$). These equations were solved using the
standard velocity Verlet integrator \cite{duenweg_1998,allen_1989},
with friction coefficient $\gamma = 0.5$ and time step $\Delta t =
0.0125$.

Both statics and dynamics of this model are known very well
\cite{kremer_1990,duenweg_1998}. In particular, its slow Rouse- or
reptation-like dynamics serves as a reference for the speedup obtained
{from} our new Monte Carlo procedure.

\subsection{Parallel Tempering}
\label{partempsec}

\begin{figure}
\centerline{\epsfig{figure=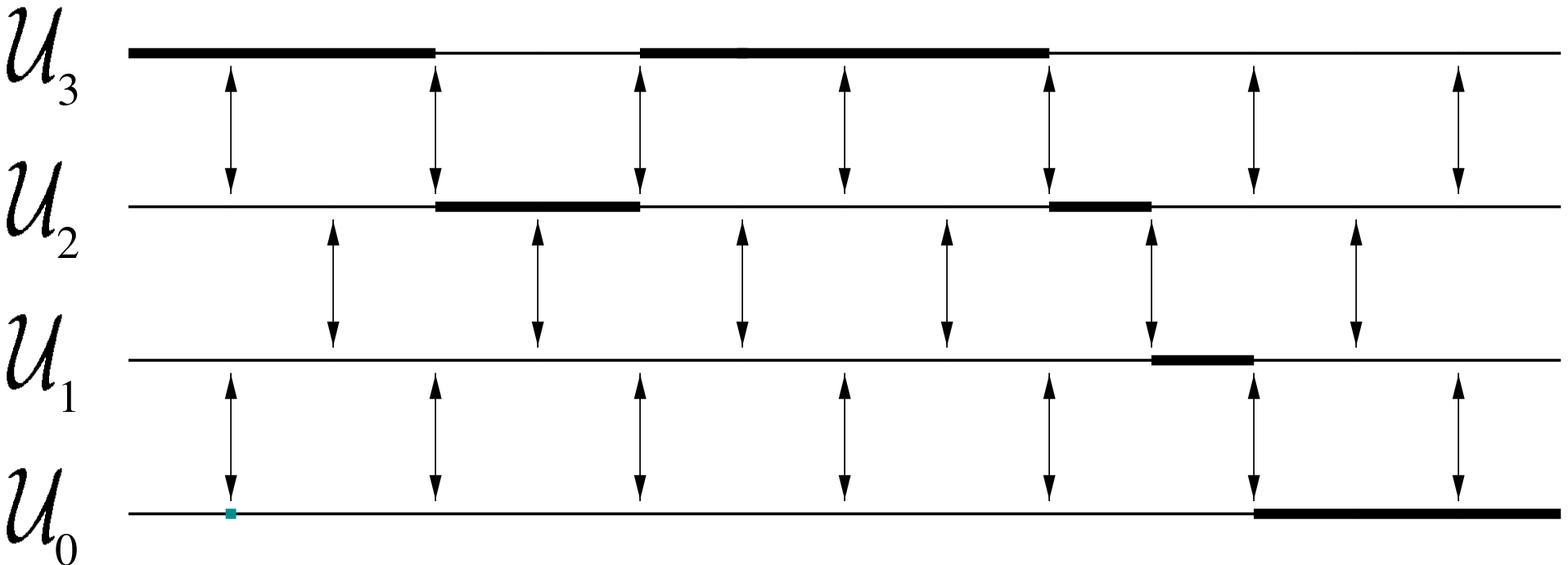,width=8.0cm}}
\vspace*{0.1in}
\caption{
  Schematic representation of our parallelization scheme. The double
  arrows represent attempted switches and the thick line represents
  the path through the Hamiltonians followed by one of the systems.  }
\label{fig2}
\end{figure}

We have performed parallel tempering
by connecting a series of systems to the Kremer-Grest potential using
successively softer repulsive core potentials. The $n$ systems are
simulated in parallel, and each system is on a separate processor of a
massively parallel system (Cray T3E). Once an initial locally equilibrated
configuration (in real and momentum space) is obtained for each of the
systems, the potentials are allowed to switch between systems through
Metropolis Monte Carlo steps, as described above. It should be noted
that the kinetic energies cancel out in the Metropolis criterion. The
swaps are implemented in a checkerboard fashion, where either the
odd-even pairs or the even-odd pairs are tried. Between these swaps
each system is run for a few stochastic dynamics steps; it is known
that this procedure is quite efficient for equilibrating local degrees
of freedom. For example, if we were to use eight processors we would
first attempt to switch the systems $1-2$, $3-4$, $5-6$, and $7-8$,
then run some stochastic dynamics, then attempt the switches $2-3$,
$4-5$, and $6-7$, then run more stochastic dynamics before attempting
the first set of switches again. Fig. \ref{fig2} shows how this
aspect of the algorithm is implemented. A reasonable duration for the
Langevin runs between the swaps is obtained from studying the
potential energy relaxation, as described later.

\begin{figure}
\centerline{\epsfig{figure=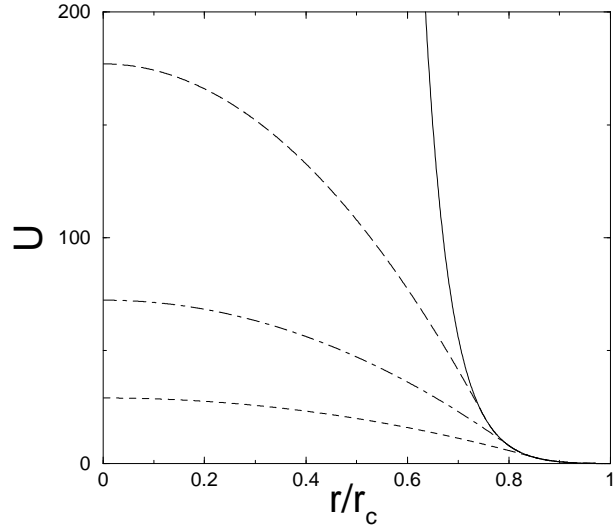,width=8.0cm}}
\vspace*{0.1in}
\caption{
  A set of successively softer repulsive core potentials. In the
  simulation we connect the system with the purely repulsive
  Lennard-Jones potential, shown as the solid line, through a series
  of such softened core repulsion potentials to the limit of phantom
  chains ($U = 0$).  }
\label{fig1}
\end{figure}

In order to achieve large acceptance rates for the swaps it is
necessary to choose the form of the ``softened core'' potential such
that the bond length $b$ and the chain stiffness $C_\infty$ are
approximately maintained. The core repulsion of the neighbors and
next-nearest neighbors on chains were thus kept intact, while all
other repulsive Lennard-Jones potentials were replaced by the
following ``softened core'' potential
\begin{eqnarray}
U_{SC}(r) &=&
A - Br^{2} \hspace*{1.23in} r \le r_t, \nonumber\\
U_{SC}(r) &=&
4\epsilon\left[\left(\frac{\sigma}{r}\right)^{12}
-\left(\frac{\sigma}{r}\right)^{6}+\frac{1}{4}\right]
\hspace*{0.2in}r_t \le r\le2^{1/6}\sigma,\nonumber\\
U_{SC}(r) &=& 0 \hspace*{1.68in} r \ge 2^{1/6}\sigma,
\end{eqnarray}
where $A$ and $B$ are fixed by the continuity of $U(r)$ and
$\frac{dU}{dr}$ leaving $r_t$ as the only free parameter. For the
Kremer-Grest potential $r_t = 0$, and $r_t$ is successively larger for
each softer potential until the final potential in the series has $r_t
= r_c = 2^{1/6} \sigma$, the cutoff radius, which is the case for phantom
chains. A graph of such a family of potentials is shown in Fig.
\ref{fig1}. We will refer to the ratio $r_t/r_c$ as the ``soft-core
parameter''. Observing Fig. \ref{fig1} it becomes quite apparent why
our tempering parameter is a superior choice to temperature for the
Kremer-Grest model. Tempering in temperature would be the equivalent
of altering the potential by a constant multiple. No matter how high
a temperature reached the $r^{-12}$ divergence of the core would not
be alleviated.

\subsection{Pivot and Translation Moves}
\label{pivotsec}

When we reach systems with extremely soft repulsive core potentials
then large scale motions within the systems will have observable
transition probabilities. We must ensure that these large scale
motions through phase space are such that only the parts of the
potential which have been softened are affected. Thus the large scale
motions should not affect any bond lengths or angles within the
polymers. Large scale moves that fulfill this criterion are pivot and
translation moves.

\begin{figure}
\centerline{\epsfig{figure=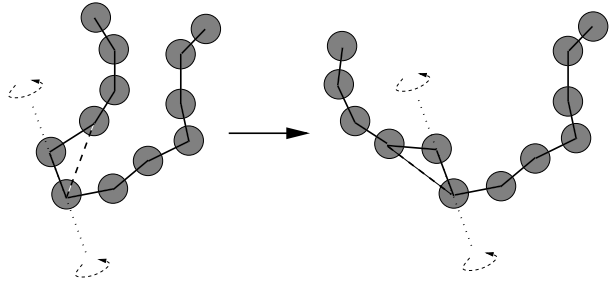,width=8.0cm}}
\vspace*{0.1in}
\caption{
  Schematic representation of our implementation of pivot moves:
  Neither the nearest neighbor nor the next-nearest neighbor distances
  on the polymer chain are affected. As a result none of the potential
  interactions which are kept at full strength are affected.  }
\label{fig3}
\end{figure}

The pivot move involves rotating part of the polymer around the axis
of a given bond. As shown in Fig. \ref{fig3}, all of the interactions
which have not been softened are unchanged in this move.  The
translation move involves taking the entire polymer and shifting it a
random distance in a random direction. It is in reaching systems where
these kinds of moves are possible before returning to the Kremer-Grest
Hamiltonian where we expect our algorithm to pay off. We have
implemented the pivot and translation moves together in a single move
where the whole chain is simultaneously translated and every bond is
rotated, thus relaxing all the degrees of freedom of the chain with
the exception of the bond lengths and angles which are relaxed by the
Langevin dynamics. We attempt these moves also quite frequently, as
discussed later in the paper.

\subsection{End-Bridging}
\label{endbridgesec}

\begin{figure}
\centerline{\epsfig{figure=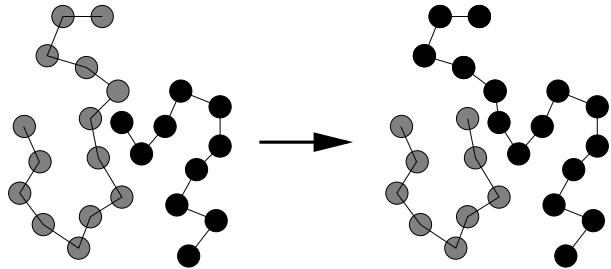,width=8.0cm}}
\vspace*{0.1in}
\caption{
  End-bridging move for the Kremer-Grest model. If the end-bridging
  move in question involves an energy change $\Delta E$ then the
  probability of the move is given by $P = \min(1,We^{-\Delta E})$
  where $W$ is a weight factor given by the ratio between the number of
  possible monomers the initial end can bridge to and the number of
  monomers the newly created end could bridge back to.  }
\label{fig4}
\end{figure}

In order to compare our algorithm with an established Monte Carlo
method for equilibrating dense polymer systems, we have also
implemented an end-bridging procedure combined with Langevin dynamics.
End-bridging, developed by Theodorou {\em et al.}
\cite{pant_1995,mavrantzas_1999} is currently the most efficient
algorithm; however, it gains its speed only by giving up
monodispersity. Instead a fixed number of monomers and a fixed number
of chains is simulated, whose length however is allowed to fluctuate
within predefined limits. In practice, these limits are defined by
allowing all chain lengths between $N (1 - f)$ and $N (1 + f)$, where
$f$ is typically of order $1/2$. The algorithm involves allowing bonds
within polymers to break and reattach to the ends of different
polymers. It was originally devised for atomistic simulations with
fixed bond lengths and involved an intricate procedure. Implementing
this algorithm on the Kremer-Grest model, however, is far simpler
since our bond lengths are able to fluctuate. The end of a chain
searches for a possible chain to bridge to. This search is performed
by finding all the monomers within the cutoff radius that are not on
the same chain as the chain end in question. One out of these is
selected at random. If a possible end-bridge is found then the move is
accepted according to a Metropolis function where the Boltzmann factor
is multiplied with a weight factor. This weight factor is given by the
number of monomers which the end could possibly bridge to divided by
the number of monomers the newly created end could bridge back to.
This is necessary to satisfy detailed balance, i.~e. to correct for
the different probabilities to select the original reaction and the
back reaction. A diagram of how end-bridging works for the
Kremer-Grest model is shown in Fig. \ref{fig4}. In our implementation
we run the Langevin dynamics for $10$ LJ time units, followed by
$n_{br}$ bridging attempts, where $n_{br}$ is $20$ times the number of
chains.

\section{Results and Discussion}
\label{resultsec}

\subsection{Construction of Simulation Procedure}
\label{construsec}

\begin{figure}
\centerline{\epsfig{figure=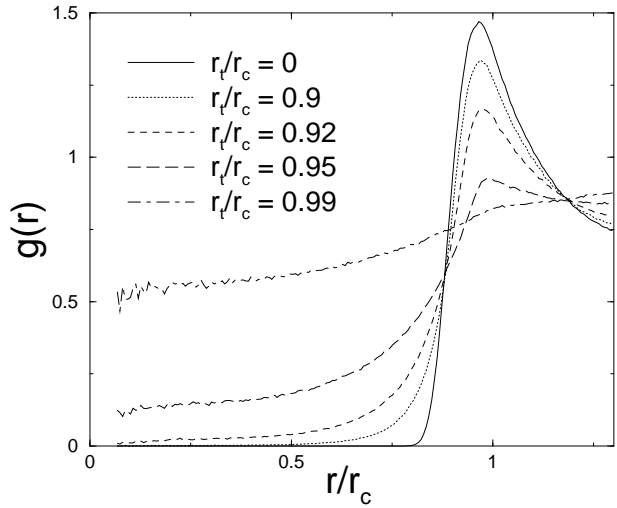,width=8.0cm}}
\vspace*{0.1in}
\caption{
  Pair correlation function for several values of $r_t/r_c$. Note that
  starting at about $r_t/r_c = 0.95$ the chains are effectively able
  to pass through each other.  }
\label{fig7}
\end{figure}

We first tested the algorithm with a system of 20 chains of length 60.
The effect of softening the potential is clearly seen in the standard
pair correlation function $g(r)$, which is the probability to find a
particle pair with distance $r$, normalized by the ideal gas value.
This function is shown in Fig. \ref{fig7}, excluding the nearest and
next-nearest neighbors on the polymer chains where the core repulsions
are maintained at full strength. Compared to the fully repulsive
system, there is a considerable probability for very short distances
as soon as $r_t / r_c \gtapprox 0.95$, reflecting the ability of
chains to pass through each other. This removes the topological
constraints for chains of arbitrary length. Thus even without pivot
moves the dynamics would not be slower than Rouse relaxation.

\begin{figure}
\centerline{\epsfig{figure=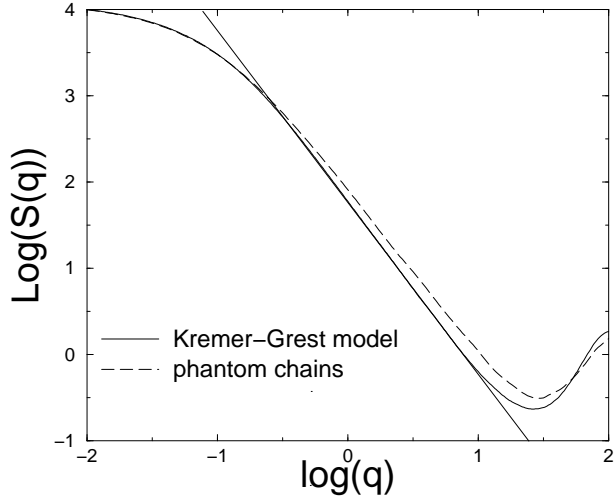,width=8.0cm}}
\vspace*{0.1in}
\caption{
  Single-chain static structure factor for both the Kremer-Grest model
  and phantom chains. The similarity in the static structure factor
  indicates that the large scale structure of the chains is nearly
  invariant as the repulsive core potentials are softened. The
  straight line represents a slope of $-2$ corresponding to $\nu =
  0.5$.  }
\label{fig8}
\end{figure}

Furthermore, we measured the single-chain static structure factor
$S(q) = N^{-1} \left< \left\vert \sum_i \exp (i \vec q \cdot \vec r_i)
  \right\vert^2 \right>$ for both the Kremer-Grest model and the
phantom chains as shown in Fig. \ref{fig8}. As expected, the Kremer-Grest
model reproduces the random walk exponent $\nu = 0.5$, observed from
the decay $S(q) \propto q^{-2}$. The result for the phantom chains is
very similar to the Kremer-Grest model, indicating that the overall
structure of the chain does not change very much as we soften the
repulsive core potentials. This in turn means that no major chain
rearrangements are necessary along the thermodynamic path, such that
the transitions should be quite easy.

\begin{figure}
\centerline{\epsfig{figure=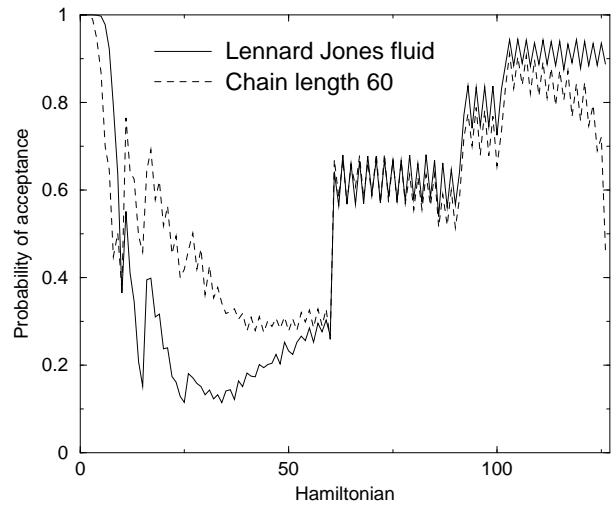,width=8.0cm}}
\vspace*{0.1in}
\caption{
  Probability that a system would transfer to the next softer
  potential, as a function of its own potential index, which
  increases with softness.}
\label{fig5}
\end{figure}

Nevertheless, it turned out that for our number of monomers one needs
of the order of $10^2$ (due to computer restrictions we used 128)
systems in order to connect from the fully repulsive potential to the
phantom chain limit, requiring that the swap acceptance rates are of
order $1/2$. The soft-core parameters were adjusted by hand, in
essence via a trial-and-error procedure. A graph of the resulting
transition probabilities for each of the potentials for both 20 chains
of length 60 and 1200 purely repulsive
Lennard-Jones particles is shown in Fig. \ref{fig5}. 
One clearly sees that in the soft limit the polymer system has much
smaller swap rates than the corresponding system of Lennard-Jones
particles. A possible explanation for this is that the chain
connectivity induces increased density fluctuations and thus an 
increased sensitivity to small changes in the interaction. We expect
this effect to become more and more pronounced with increasing chain length.
This is a severe problem
for this algorithm for two reasons. The algorithm will be slowed
by an extra factor above and beyond that determined in our initial
scaling analysis with increasing chain length and 
any pre-determination of an appropriate set of $r_t$ values
based on a simple model with transferability to more complicated
systems is precluded. In any implementation
of this algorithm to a new problem the set of $r_t$ values
for the specific model will have to be determined.
We have not devised a systematic method for
selection of the soft-core parameters. This would be a future
improvement of the algorithm.

\begin{figure}
\centerline{\epsfig{figure=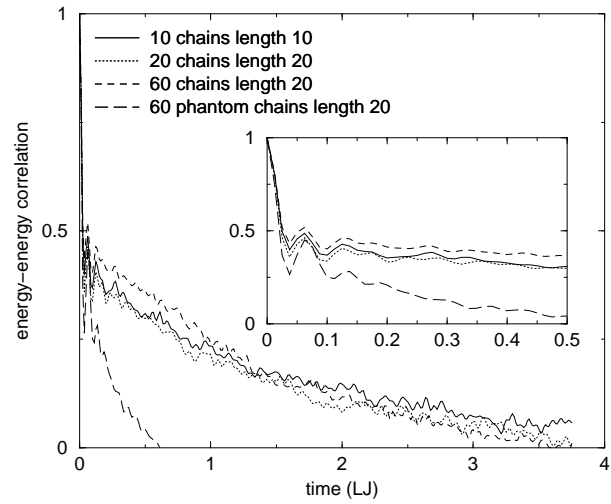,width=8.0cm}}
\vspace*{0.1in}
\caption{
  Energy-energy correlation function as a function of time in
  Lennard-Jones time units. Note the very fast decay on short time
  scales followed by a much slower decay which is dependent on the
  soft-core parameter.}
\label{fig6}
\end{figure}

{From} measuring the time autocorrelation function of the potential
energy for the untempered system (i.~e. without swaps between
potentials), shown in Fig. \ref{fig6}, we have found that the
correlations decay very quickly after only a few time steps of
Langevin dynamics and then cross over to a much slower decay. Our
interpretation of the very fast initial decay is that it is a direct
consequence of the local bond oscillations, which happen on roughly
this time scale. 
 This is also in accord with the observation that it
occurs independently of the degree of softening, see Fig. \ref{fig6},
since the potentials between bond neighbors remained
unchanged for all potentials. Conversely, the long-time behavior is quite strongly
affected by the softening, again in agreement with the expectation
that the function should decay much faster for a softer system.
Furthermore, the energy-energy autocorrelation function is independent
of chain length, as expected.

These results suggest that it is most efficient to attempt the
switches frequently, on the time scale of the bond oscillations.  We
thus constructed the following algorithm: On every system we perform
the large scale chain reorientation attempt on enough chains so that
attempts are made to move at least 5\% of the monomers. This is
followed by 4 Langevin dynamics steps, after which the Hamiltonian
swaps (either even-odd or odd-even) are performed.  Then the procedure
is repeated. For reasons of simplicity, we have applied the identical
moves to all systems. It should be stated, in the interest of future
development of this method, that this condition is however not
necessary. Any algorithm
which leaves the Boltzmann distributions of the different systems
invariant will be valid. For our algorithm, however,
as currently implemented, not to attempt the large-scale moves
on the hard systems would only generate idle CPU time since the
processors all have to wait for the slowest system to finish
before attempting the next swap.  A further
optimization could however involve eliminating the large-scale moves
on the hard systems, and replacing them by more Langevin steps. By
fine-tuning this one should be able to reduce synchronization overhead
to a minimum. We also found that as the soft-core parameter approaches
the phantom chain limit local bond oscillations can become unstable.
This can be easily remedied by increasing the friction in the
Langevin dynamics as the soft-core parameter increases.

\subsection{Correlation Functions}
\label{corrfuncsec}

An appropriate way to benchmark the program is to determine the CPU
time needed per chain relaxed. Since every system periodically passes
through the Hamiltonian with the full Lennard-Jones repulsive hard
core potential, each system can be seen as a Kremer-Grest model which
is sampled every time this Hamiltonian happens to lie on it. Useful
single-chain quantities to measure are the normalized end-to-end vector
autocorrelation function, $\left< \vec R (t) \cdot \vec R(0) \right>$
with $\vec R = \vec r_N - \vec r_1$ and the autocorrelation function of the lowest five Rouse
modes, $\left< \vec X_p (t) \cdot \vec X_p (0) \right>$ with $p = 1,
\ldots, 5$ and \cite{kopf_1997}
\begin{equation}
\vec X_p  =  \sqrt{2} N^{-1/2} \sum_{i=1}^N \vec r_i
        \cos \left[ \frac{p \pi}{N} \left( i - 1/2 \right) \right] .
\end{equation}
These quantities must be measured in such a way that only correlations
between configurations where the Kremer-Grest Hamiltonian is present
are counted. Thus we define the following procedure to measure
autocorrelation functions:
\begin{equation}
C(t) = \frac{
\sum_{t'} u(t' + t) u(t') \delta(U(t' + t),U_0) \delta(U(t'),U_0)}{
\sum_{t'} \delta(U(t' + t),U_0) \delta(U(t'),U_0)} ,
\end{equation}
where $u(t)$ is the quantity whose autocorrelation function is
determined, $U_0$ is the Kremer-Grest potential, and $U(t)$ the
potential at time $t$. The Kronecker $\delta(U(t),U_0)$ vanishes
unless the potential is $U_0$ where it is unity.

\begin{figure}
\centerline{\epsfig{figure=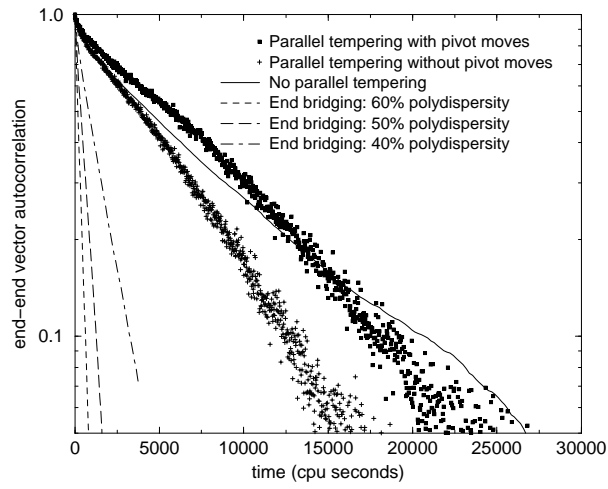,width=8.0cm}}
\vspace*{0.1in}
\caption{
  Normalized end-to-end vector autocorrelation function for standard
  Langevin dynamics, our parallel tempering procedure with and without 
  pivot moves, and end-bridging simulations for various degrees of
  polydispersity.}
\label{figendend}
\end{figure}

\begin{figure}
\centerline{\epsfig{figure=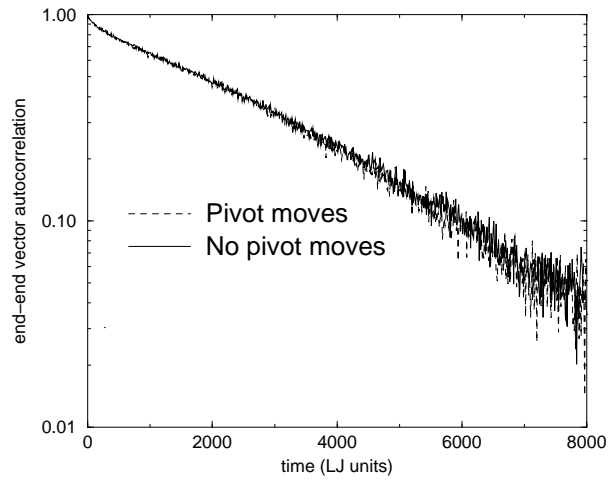,width=8.0cm}}
\vspace*{0.1in}
\caption{
Normalized end-to-end vector autocorrelation function 
for our parallel tempering
algorithm with and without pivot moves.
}
\label{figurepivotendend}  
\end{figure}

\begin{figure}
\centerline{\epsfig{figure=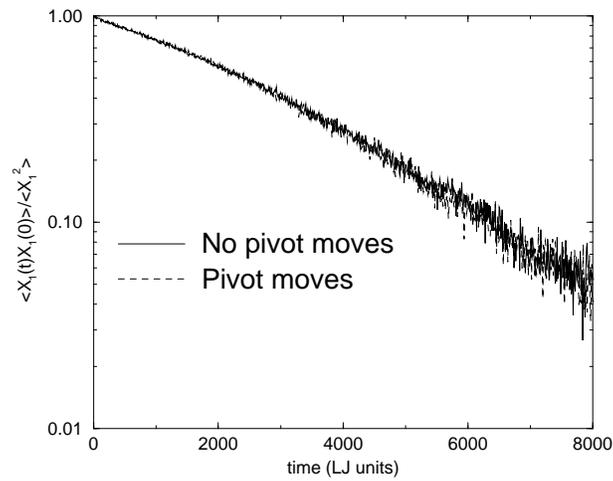,width=8.0cm}}
\vspace*{0.1in}
\caption{
Autocorrelation of the first Rouse mode for our parallel tempering
algorithm with and without pivot moves.
}
\label{figurepivotrouse}
\end{figure}

Fig. \ref{figendend} displays the normalized end-to-end vector
autocorrelation function for (i) end-bridging with different degrees of
polydispersity $f = 0.4 \ldots 0.6$, (ii) our simulated tempering
procedure, and (iii) standard Langevin (Rouse-like) dynamics. It is
seen that the end-bridging procedure is by far the fastest; however,
it should be kept in mind that the speed comes at the price of
considerable polydispersity. Furthermore, the end-to-end vector does
not describe the slowest relaxation in the system for this type of
algorithm. Since for the end-bridging algorithm what constitutes a
polymer chain becomes an ill defined quantity, what actually does
become the slowest mode is unclear.  The simulated tempering procedure
turns out to be somewhat faster than plain Rouse relaxation, in
particular in the long-time limit. In terms of integrated
autocorrelation time the speedup amounts to roughly 10 \% with the
pivot moves present and 30\% without. The reason why pivoting actually
slows the simulation down is explained from Figs.
\ref{figurepivotendend} and \ref{figurepivotrouse}, which show the
normalized end-to-end vector autocorrelation function and the
autocorrelation of the first Rouse mode (in terms of ``physical''
time, not CPU time). The simple parallel tempering already needs such
a long time for traversing Hamiltonian space from the ideal gas to the
full repulsive interaction that on this time scale the chains are
already fully relaxed. After a ``diffusive loop'' through Hamiltonian
space the configuration is thus already fully decorrelated, even
without pivot moves. Therefore the pivot moves just generate
additional CPU overhead and cause a slowdown.

\begin{figure}
\centerline{\epsfig{figure=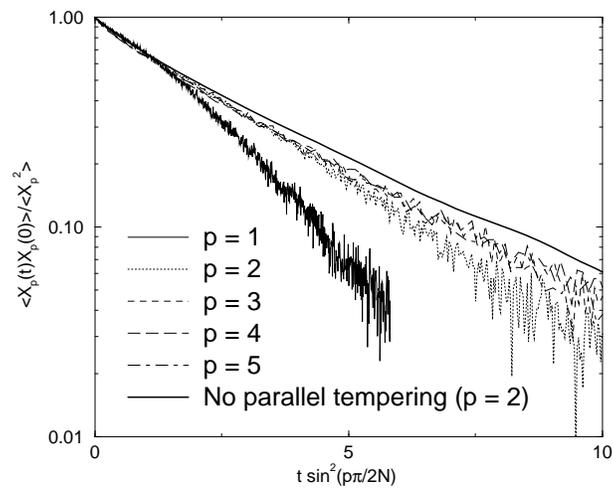,width=8.0cm}}
\vspace*{0.1in}
\caption{
Rouse mode analysis for the lowest five Rouse modes for
our parallel tempering procedure and compared to the
case of pure Langevin dynamics. Though we have shown
only the result for the second Rouse mode, all Rouse modes
fall onto the same line for pure Langevin dynamics.
}
\label{rousemodes}
\end{figure}

In Fig. \ref{rousemodes} the autocorrelation of the lowest five Rouse
modes vs. $t\sin^2(p\pi/2N)$ is shown. If Rouse scaling holds, then
all Rouse modes should collapse onto a single line. As is known from
older simulations \cite{kremer_1990,kopf_1997}, $N = 60$ is already
slightly in the crossover regime to reptation, where ultimately the
lower modes are slowed down. Nevertheless, $N = 60$ is still too short
for this effect to be become visible, such that Rouse behavior for the
case of pure Langevin dynamics can still be assumed. Our tempering
procedure, on the other hand, produces a disproportionate acceleration
of the lower modes. They first decay exponentially in accord with the
pure Rouse dynamics of the hard system. At later times, however, one
observes a steeper exponential decay, which is the effect of the
decorrelating excursions to the softer systems.

\begin{figure}
\centerline{\epsfig{figure=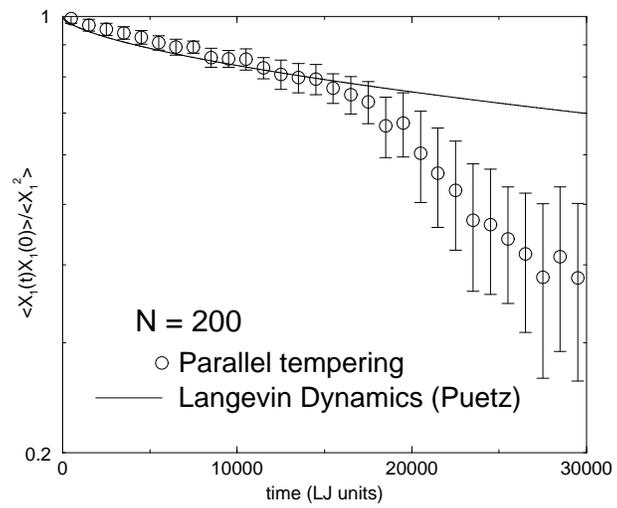,width=8.0cm}}
\vspace*{0.1in}
\caption{
  Normalized autocorrelation of the first Rouse mode for 
  our parallel tempering procedure 
  for chain length $200$. We have compared our result to a fit to
  results previously obtained by M. P\"{u}tz{\protect \cite{mattias}}. 
  }
\label{fig200}
\end{figure}

We were also able to obtain results, shown in Fig. \ref{fig200}, for
a system of 32 chains of length 200, using 256 Hamiltonians.
Limitations in the CPU time available to us have prevented us from
performing a comparison between simulations with and without pivot
moves, and from measuring the correlation function until full decay.
Nevertheless, our preliminary data clearly show a steep drop off in
the correlation functions, which is much more pronounced than for the
case of $N = 60$. These data were obtained for the case {\em with} pivot
moves; we believe that they actually did help to accelerate the
equilibration of this system. The data for pure Langevin dynamics
without tempering were taken from Ref. \cite{mattias}. Extrapolating
our data, we guess that the speedup is a factor of five or
even larger.

\subsection{Conclusions}
\label{conclusec}

Our results indicate that parallel tempering combined with large-scale
chain moves is, in principle, a viable route to speeding up
simulations of dense polymer systems. However, for chains of moderate
length ($N = 60$) the effort seems hardly worthwhile, since the
increase in efficiency compared to simple Langevin dynamics is not
spectacular. Nevertheless, it is expected that the method will pay off
for systems with either longer chains, as indicated from our
preliminary results for chain length $200$, or more complicated
molecular architecture. These issues will be the subject of future
investigations. Current trends in the development of computational
facilities indicate that over the next decade we will see an increase
in the availability of massively parallel computers with more and more
processors running at approximately the speed of today's processors.
With the advent of such facilities we expect the full potential of
this algorithm to be realized.

\section*{Acknowledgments}

We thank J. J. de Pablo, R. Everaers, and A. Khokhlov for helpful
suggestions and stimulating discussions. A. B. thanks D. Theodorou and
V. I. Mavrantzas for hospitality at the Institute of Chemical
Engineering and High Temperature Chemical Processes, FORTH, Patras,
Greece, where part of this work was done. This research was supported
by the EU TMR network ``NEWRUP'', contract ERB-FMRX-CT98-0176.

\clearpage

\clearpage

\clearpage

\clearpage

\clearpage

\clearpage

\clearpage

\end{document}